# Authentic Digital Signature Based on Quantum Correlation

Xiao-Jun Wen, Yun Liu

School of Electronic Information Engineering, Beijing Jiaotong University, Beijing 100044, China

**Abstract:** An authentic digital signature scheme based on the correlation of Greenberger-Horne-Zeilinger (*GHZ*) states was presented. In this scheme, by performing a local unitary operation on the third particles of each *GHZ* triplet, Alice can encode message *M* and get its signature $|S\rangle$. Bob performs $C_{NOT}$ operation on the combined *GHZ* triplets can recovery the message *M* and directly authenticates Alice's signature $|S\rangle$. Our scheme was designed to use quantum secret key to guaranteed unconditional security, as well as use quantum fingerprinting to avoid trick attacks and reduce communication complexity.

## 1. Introduction

According to the mode of authenticating signature, the digital signature protocols are divided into two categories: (1) direct authenticate signature technique, (2) indirect authenticate signature technique. In the direct check signature technique, the receiver executes authenticating signature directly, so it is called authentic digital signature for short. However, in the indirect authenticate signature technique, authenticating signature is executed by the third party who serves as an arbitrator, so it is called arbitral digital signature [1].

In an arbitrated signature scheme, the arbitrator such as system manager can have access to the contents of the messages; therefore, the security of most arbitrated signature schemes depends heavily on the trustworthiness of the arbitrators. And especially, the arbitrator who participates in authenticating signature directly may reduce the efficiency of the whole system. On the contrary, the authentic digital signature has no these problems, but in fact, though the direct authenticate signature technique need not an arbitrator to authenticate signature, the issuer and verifier may ask the third party for arbitration if some disputes had happened between them.

As classical cryptography, quantum secrecy communication involves the problem of quantum information signature. Recently, the research of quantum communication and quantum computation is progressing rapidly. The experiment of Quantum Key Distribution (*QKD*) [2][3] and Quantum Secret Sharing (*QSS*) [4][5] had achieved success. Quantum Message Authentication (*QMA*) [6] and Quantum Signature (*QS*) [7] are attracted by academe. In the classical cryptography, many digital signature schemes based on computational security, such as ElGamal and Digital Signature Algorithm (*DSA*), however, the quantum computer will break these schemes easily.

Different from the classical cryptography, the quantum information technology is based on physical principles, so quantum signature can have unconditional security. Gottesman and Chuang proposed a quantum digital signature scheme [7] based on quantum one-way function, and claimed that their scheme was secure against quantum attack. However, each classical bit of their scheme needs to be distinguished two outputs $|f_k\rangle$ and $|f_{k'}\rangle$ for authenticating its signature, that is very cumbersome and inefficient as they wrote in the paper. Zeng[8][9] and Lee[10] had researched the quantum information signature schemes which were based on the correlation of the *GHZ* triplet states[11], but their schemes belong to the arbitrated signature schemes which have many disadvantages as above described. We [12] had proposed a signature scheme based on Quantum Entanglement Swapping (*QES*) before, however, quantum entanglement swapping needs more particles to participate in, so the communication efficiency is lower in this scheme.

In this paper, we proposed a novel quantum digital signature scheme based on the correlations of the *GHZ* states and the character of local unitary operation. The features of our scheme include three aspects: 1) it has the function of message recovery, and the signature information is hided in some particles, instead of appending a special attachment to the message 2) it does not need an arbitrator to authenticate the signature unless some disputes had happened between the issuer and verifier 3) it has provable unconditional security and use quantum fingerprinting function to reduce the communication complexity.

The paper was outlined as follows. In section 2, we introduced the basic theory that how the *GHZ* states and local unitary operation can be applied in quantum signature. In section 3, we proposed the signature and authentication scheme based on the quantum correlation. The security was analyzed and we discussed the results in section 4. In section 5 we presented some conclusions.

## 2. Basic theory

Define the four Bell states of 2-qubit as

$$|\Phi^+\rangle = \frac{1}{\sqrt{2}}(|00\rangle + |11\rangle) \qquad (1)$$

$$|\Phi^-\rangle = \frac{1}{\sqrt{2}}(|00\rangle - |11\rangle) \qquad (2)$$

$$|\psi^+\rangle = \frac{1}{\sqrt{2}}(|01\rangle + |10\rangle) \qquad (3)$$

$$|\psi^-\rangle = \frac{1}{\sqrt{2}}(|01\rangle - |10\rangle) \qquad (4)$$

The *GHZ* state is an entangled state of a three qubits system, which is expressed as:

$$|\psi\rangle = \frac{1}{\sqrt{2}}(|000\rangle + |111\rangle) \qquad (5)$$

Now, we perform two style transforms on above state as follows:

Transform 1: If perform a $C_{NOT}$ operation on the first two qubits (the first qubit as the target qubit and the second qubit as the control qubit), the state of the tripartite system is transformed into:

$$|\pi\rangle = \frac{1}{\sqrt{2}}(|000\rangle + |011\rangle)$$
$$= |0\rangle \otimes \frac{1}{\sqrt{2}}(|00\rangle + |11\rangle) \qquad (6)$$

The tripartite system has been divided into two independent subsystem and the last two qubits just in a Bell state $|\Phi^+\rangle$.

Transform 2, If perform a unitary operation $\hat{\sigma}_x$

$$\hat{\sigma}_x = \begin{pmatrix} 0 & 1 \\ 1 & 0 \end{pmatrix} \qquad (7)$$

on the third qubit of the *GHZ* state,

We get

$$|\psi'\rangle = \frac{1}{\sqrt{2}}(|001\rangle + |110\rangle) \qquad (8)$$

Performing the $C_{NOT}$ operation on above state as transform 1, we get:

$$|\pi'\rangle = \frac{1}{\sqrt{2}}(|001\rangle + |010\rangle)$$
$$= |0\rangle \otimes \frac{1}{\sqrt{2}}(|01\rangle + |10\rangle) \qquad (9)$$

The subsystem of the last two qubits is also in another Bell state $|\Psi^+\rangle$.

So we can judge whether we had performed unitary operation $\hat{\sigma}_x$ on the *GHZ* state or not by doing the Bell-state measurement on the last two qubits, this theory can be used to realize message recovery, signature and authentication.

## 3. Signature scheme description

In this signature scheme, Alice wants to send a digital message *M* and its signature |*S*⟩ to Bob, Bob will authenticate the signature once he receives the message and its signature. If the signature is trustful, Bob will accept the message, otherwise he'll reject it, and they may ask the system manager Trent for arbitration in case of some disputes.

*(1) Initial Phase*

1) Key distribution. Alice and Bob share a secret key *K* by using quantum key distribution (*QKD*) protocols [2] such as BB84 proved as unconditionally secure [3,13].

2) *GHZ* triplets' distribution. Before Alice sends a signature message to Bob, Bob prepares *N GHZ* triplets, each triplet in the state |*ψ*⟩$_{123}$, we denote these *N*-triplet as {|*ψ*(1)⟩$_{123}$, |*ψ*(2)⟩$_{123}$,..., |*ψ*(*N*)⟩$_{123}$ };

3) Alice transforms the message *M* into *N*-bit as {*M*(1), *M*(2),…, *M*(*N*)}, and they consult with that "1" corresponds to performing $\hat{\sigma}_x$ operation and "0" corresponds to no any operation respectively. For example, *M* = {011011} corresponds to the operations "×$\sigma_x\sigma_x$×$\sigma_x\sigma_x$"("×" denotes doing nothing on the corresponding particles).

4) To guarantee the integrity of the message *M*, Alice choices a quantum one-way function |*f*(*x*)⟩, which is easy to calculate |*f*(*M*)⟩ but is impossible to invert *M* from |*f*(*M*)⟩ by virtue of fundamental quantum information theory. Here, we chooses the quantum fingerprinting [14] as the quantum one-way function (with a classical input and a quantum output).The quantum fingerprinting function of *n*-bit inputs *x* was defined as

$$|f(x)\rangle = \frac{1}{\sqrt{m}}\sum_{i=1}^{m}|i\rangle|E_i(x)\rangle \quad (10)$$

Where $x \in \{0,1\}^n$, $E:\{0,1\}^n \to \{0,1\}^m$ is a family of error correcting code with fixed *c* > 1, and *m* = *cn*, |*f*(*x*)⟩ is a (log(*m*) + 1)-qbit quantum states. Since |*f*(*x*)⟩ changes according to the value of parameter *c*, Alice secretly chooses *c* as her private key.

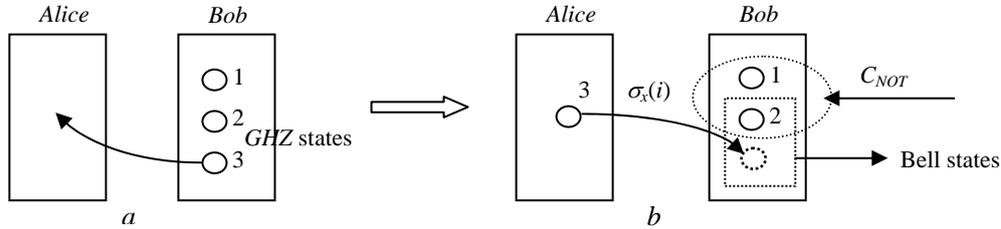

Figure 1. Schematic of transmission, unitary operation and Bell measurement on the *i*th triplet (a. Bob sends particle 3 to Alice b. Alice performs $\sigma_x(i)$ operation to particle 3,than sends it back to Bob. Bob performs $C_{NOT}$ operation on the first two qubits, then does Bell-state measurement on the last two qbits.)

*(2) Signature Phase*

1）Bob sends the third qubits of each triplet to Alice (see figure 1. a.), their states are expressed as {|*ψ*(1)⟩$_3$,|*ψ*(2)⟩$_3$,…, |*ψ*(*N*)⟩$_3$};

2）After having received these particles 3, Alice performs $\sigma_x(i)$ operations on the *i*th particles according to the code rule of *M*(*i*)(*i* = 1,2,…,*N*), the states of these transformed particles are expressed as

$$|S\rangle = \{|\psi'(1)\rangle_3, |\psi'(2)\rangle_3, ..., |\psi'(N)\rangle_3\} \quad (11)$$

Where |*S*⟩ is the signature generated by Alice, that is to say, the transformed particles include the information of Alice's signature;

3）Alice sends these transformed particles 3 ( $|S\rangle$ ) back to Bob(see figure 1. b.);

4) To prevent *M* from being forged by attackers, Alice encrypts the message *M* to get $E_K\{M\}$ using her secret key *K*;

5）Alice sends $E_K\{M\}$ to Bob.

*(3) Authentication Phase*

Bob can authenticate the Alice's signature directly in the following steps:

1）After receiving particles 3, Bob combines them with the corresponding two qubits in her hand, we record the states of them as $\{|\psi'(1)\rangle_{123}, |\psi'(2)\rangle_{123}, ..., |\psi'(N)\rangle_{123}\}$;

2）To every combined triplet, Bob performs $C_{NOT}$ operation on the first two qubits, then the states of the new triplets are expressed as $\{|\pi(1)\rangle_{123}, |\pi(2)\rangle_{123}, ..., |\pi(N)\rangle_{123}\}$;

3）Bob does Bell state measurement on the last two qbits of each triplet, if the measurement outcome is $|\Phi^+\rangle$, he writes down "0" and when it is $|\Psi^+\rangle$, he writes down "1". At last, Bob will get an *N*-bit string *M'*;

4）Bob decrypts $E_K\{M\}$ to get *M* by the secret *K* shared with Alice;

5）Bob compares *M* and *M'*, if *M* = *M'*, then accepts $|S\rangle$ as the truthful signature of message *M*.

If both Alice and Bob are trusted, then the process of signature and authentication may end by now, else it will turn into the following arbitration phase.

*(4) Arbitration Phase*

(1)Before the end of once signature and authentication process, Alice transforms the message *M* into $|f(M)\rangle$ by the quantum fingerprinting function with the secret value of the parameter *c*(as her private key), then Alice sends $|f(M)\rangle$ (as the public key) to the arbitrator Trent, we denotes it as $|f(M)\rangle_T$.

(2) If some disputes have happened, Alice reproduces a $|f(M)\rangle$ according to his message *M*, we denotes it as $|f(M)\rangle_A$. Trent compares $|f(M)\rangle_T$ and $|f(M)\rangle_A$, if $|f(M)\rangle_T = |f(M)\rangle_A$, then judges Alice's signature $|S\rangle$ is valid.

The comparison of two quantum states $|f(M)\rangle_T$ and $|f(M)\rangle_A$ is less straightforward than in the classical case because of the statistical properties of quantum measurements. Here, we use the Quantum Swap Test Circuit (*QSTC*) [14] which was proposed by Harry Buhrman and his co-workers to compare whether $|f(M)\rangle_T$ and $|f(M)\rangle_A$ are equivalent or not.

# 4．Security analysis and discuss

*(1)The quantum attack strategy cannot be successful*

Eve adopts capture/resend strategy. Suppose that Eve, one of adversaries who know this signature protocol very well, captures the qubits which Alice sends to Bob, he will performs $\hat{\sigma}_x$ or other unitary operations on these particles in order to tamper message *M* and its signature $|S\rangle$.

However, Eve's attack strategy can not be successful in this scheme. Because the message *M* is encrypted by the quantum secret key *K* shared between Alice and Bob at the initial phase. Owing to unconditional security of a quantum key distribution Eve cannot detect this key. Hence it is impossible that Eve tampers the message *M* and counterfeits Alice to sign the message.

*(2) The classical trick strategy cannot be successful*

1) The message receiver can not forge signature

Suppose that the message receiver Bob is not honest, he attempts to forge Alice's signature in order to get profit for himself. Because all qubits in each *GHZ* triplet are in his hand at the end of the scheme, so he has chances to forge $|S\rangle$ signed by Alice. Even more terrible, Bob knows the key *K* shared with Alice, thus he can counterfeit Alice to generate $E_K\{M\}$ and $|S\rangle$.

If this dispute happens, Alice may ask Trent for arbitration. Trent compares $|f(M)\rangle_T$ and $|f(M)\rangle_A$ would judge whether Bob had counterfeited Alice to generate $E_K\{M\}$ and $|S\rangle$. For $|f(M)\rangle_A$ includes the information of Alice's private key *c*, it cannot be forged by others, therefore, Bob's trick cannot be successful.

2) The message sender can not disavow his signature

In this scheme, Alice should send $E_K\{M\}$ except for the signature $|S\rangle$ to Bob. Since $E_K\{M\}$ includes the key $K$ which is known only by Alice and Bob, So Alice cannot disavow her signature in general. However, Alice may intentionally say Bob also knows the key $K$ so as to disavow her signature.

If this dispute happens, Bob may ask Trent for arbitration. Since $|f(M)\rangle_T$ includes Alice's private key $c$, so Alice cannot disavow her signature.

*(3) Comparison about communication complexity with other quantum signature scheme*

Compared to **[8-10]** whose signature schemes are based on *GHZ* states, our scheme does not require an arbitrator to authenticate the signature unless some disputes happen, and the fingerprints which Alice sends to the arbitrator are exponentially shorter than in the comparable classical setting[14]. As stated in Table 1, we have improved the communication efficiency in this scheme.

**Table1:** The quantity of transmitted qbits for the quantum signature schemes (QSS) when using the *n*-bit message

| Transmission of the signature | QSS in [8,9] | QSS with a public board in [10] | QSS without a public board in [10] | Our QSS |
|---|---|---|---|---|
| Alice → Bob | 3n | 2n | 2n | n |
| Bob → the arbitrator | 3n | 3n | 3n | - |
| The arbitrator → Bob | 5n + 1 | 3n + 2 | 4n + 1 | - |
| Bob → Alice | - | - | - | n |
| Alice → the arbitrator | - | - | - | log(cn) + 1 |

## 5. Conclusions

In this paper, an authentic digital signature scheme based on correlation of *GHZ* states was presented. In this scheme, Alice performs a local unitary operation on the third particles of each *GHZ* triplet according to the message code, these particles including Alice's signature information are sent to Bob. Bob performs $C_{NOT}$ operation on the combined *GHZ* triplets can recovery the message *M* and authenticates Alice's signature $|S\rangle$ according to the correlations of *GHZ* states.

To prevent attackers from forging message *M* or signature $|S\rangle$, our scheme used *QKD* and quantum one-time pad algorithm to guarantee their security, furthermore, we designed quantum fingerprinting to avoid tricks from dishonest Alice and Bob. Our scheme provided unconditional security and high communication efficiency, realization of this scheme is completely possible according to current technology and experiment.

## 6. Acknowledgments

This work was supported by the National Natural Science Foundation of China, Grants No: 60572035; also supported by the Foundation of Beijing Municipality Key Laboratory of Communication and Information System (No. JD100040513).

**Reference:**
[1] William Stallings, *Cryptography and network security:principles and practice:Second Edition*, Prentice Hall, 2003 pp.299-305
[2] Hughes R, *et al*,Quantum key distribution over a 48km optical fibre network, **J.Mod.Opt** ,2000,**47**,p.533
[3] Shor P, *et al*.: *Simple* Proof of Security of the BB84 Quantum Key Distribution Protocol, **Phys.Rev. Lett.** 2000,**85**,p.441
[4] Hillery M, *Buzek* V, Berthiaume A. Quantum secret sharing. **Phyical Review A**,1999,**59**(3):1829-1834.
[5] Guo G, *et al*. Quantum secret sharing without entanglement. **Phys .Lett. A,**2003, **310**(4):247-251.
[6] Barnum C, *et al*..Authentication of quantum messages[A].**Proceedings of 43rd Annual IEEE Symposium on the**

**Foundations of Computer Science**,Vancouver,2002,449-458.

[7] Gottesman D, Chuang I. Quantum digital signatures, [2001-05-08].*http://arxiv.org/abs/quant-ph/0105032.pdf*

[8] Zeng G, et al. Signature scheme based on quantum cryptography, **Acta Electronica Sinica (in Chinese)**, 2001，**29**(8):1098-1100

[9] Zeng G, Christoph K. An arbitrated quantum signature scheme[J].**Physical review A**. 2002, **65**(4): 042312-042317

[10] Lee H, Hong C, Kim H, *et al.* Arbitrated quantum signature scheme with message recovery. **Physics Letters A**,2004,**321**: 295-300.

[11] Greenburger D, Horne M, *et al*, Bell's theorem without inequalities, **Am. J. Phys.**, 1990,**58**: 1131-1143.

[12] Wen.X, *et al.* Signature Scheme Based on Qantum Entanglement Swapping **journal of electronics & information technology (in chinese)**, 2005,**27**(5),811-813

[13] Mayers D, Unconditional security in quantum cryptography,**Journal of the ACM**,2001,**48**(3): 351-406

[14] Buhrman H, et al.Quantum fingerprinting, **Phys.Rev. Lett**, 2001, **87**:167902-167904